\documentclass{aa}
\newcommand{\kms}{km\,s$^{-1}$}
\newcommand{\Msun}{M_\odot}
\newcommand{\grad}{^\circ}
\usepackage{graphicx,wrapfig,amssymb}
\def\ltsimeq{\,\raise 0.3 ex\hbox{$ < $}\kern -0.8 em \lower
0.7 ex\hbox{$\sim$}\,}
\begin{document}
\title{Q0957+561 revised: CO emission from a disk at $z$ =
  1.4\thanks{Based on observations carried out with the IRAM Plateau
    de Bure Interferometer. IRAM is supported by INSU/CNRS (France),
    MPG (Germany) and IGN (Spain).}}


   \author{M.\ Krips
          \inst{1}\fnmsep\inst{2}
      \and
          R.\ Neri\inst{2},
          A.\ Eckart\inst{1},
      D.\ Downes\inst{2},
      J.\ Mart\'{\i}n-Pintado\inst{3}
      \and
      P.\ Planesas\inst{4}
          }
   \offprints{\tt krips@ph1.uni-koeln.de}

   \institute{I. Physikalisches Institut, University of Cologne,
              Z\"ulpicherstr. 77, 50937 K\"oln, Germany
         \and
             Institut de Radio Astronomie Millim\'etrique,
             300 rue de la Piscine, 38406 Saint Martin d'H\`eres, France
     \and
         Instituto de Estructura de la Materia (CSIC),
         Serrano 121, 28006 Madrid, Spain
     \and
             Observatorio Astron\'omico Nacional (IGN),
         Apartado 112, 28800 Alcal\'a de Henares, Spain
}
   \date{Received }

   \abstract{Based on additional interferometric observations, we
     reanalysed the CO(2-1) and 3mm continuum emission of Q0957+561, a
     lensed QSO at a redshift of $z=1.4141$. The emission in the
     CO(2-1) lines reveals a gas-rich host galaxy with a peculiar
     double-peaked profile at one of the two lensed images. Our new
     interferometric CO maps of the host galaxy agree well with HST
     images obtained by Keeton et al.\ (2000) and we thus argue that
     the two velocity components arise from molecular gas in the disk
     of the host galaxy. We also present new model calculations, all
     in excellent agreement with recent time delay measurements and
     simulations. 
   }
   \maketitle
%

\section{Introduction}

Since the discovery of Q0957+561, the first confirmed gravitationally
lensed quasar (Walsh et al.\ \cite{wals}) at a redshift of $z=1.4141$,
several models have been developed to understand the lensing potential
of the interve\-ning ga\-laxies: a giant elliptical galaxy (G1) at a
redshift of $z = 0.355$ with a surrounding cluster and probably
another group of background galaxies at $z = 0.5$ (Angonin-Willaime et
al.\ \cite{ango}; Chartas et al.\ \cite{char}). Besides flux
amplification, gravitational lensing helps also to improve spatial
resolution as detailed by Kneib et al.\ (\cite{knei}). Differential
lensing effects which are detectable through the comparison of lensed
images at different wavelengths indeed provide a powerful tool to
probe the structure of a galaxy at a much higher resolution than
possible with current millimeter interferometers. Planesas et al.\
(\cite{plan}, in the following P99) have succeeded in detecting the
CO(2--1) line in the host galaxy around Q0957+561, but lacked a
detailed lensing model to investigate the source of mo\-le\-cu\-lar
emission. Keeton et al.\ (2000) have recently detected also the
stellar component of the host galaxy showing differences in its
distribution with respect to those of the gas obtained by P99.

To confirm and substantiate the weak line profiles previously detected
by P99 and to eliminate the previous inconsistencies to Keeton et al.\
(2000), we have carried out new observations with the IRAM
interferometer. To further improve on P99's interpretation, we have
developed a numerical code which incorporates existing lensing models
of Q0957+561\footnote{We adopt a flat cosmology with $\Omega_M$=0.3,
$\Omega_V$=0.7 and a Hubble constant of 65 \kms Mpc$^{-1}$.}.

\begin{figure*}[!]
  \centering 
  \resizebox{\hsize}{!}{\rotatebox{-90}{\includegraphics{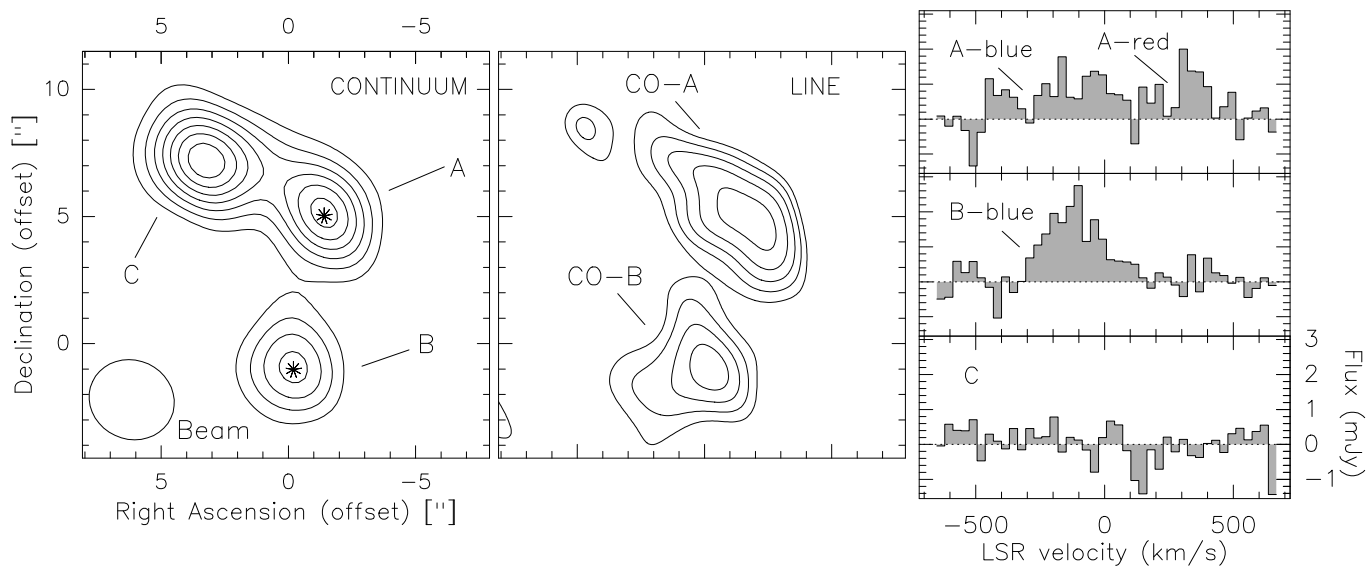}}}
  \caption{Continuum at 3mm ({\it left panel}), integrated CO(2--1)
  line emission ({\it middle panel}) and the spectral CO(2--1)
  profiles observed toward CO$-$A, CO$-$B and at the 3mm position of
  the radio jet ({\it right panel}). Offset position (0,0) is at
  the assumed position of the lens, i.e.\ at $\alpha_{\rm
  J2000}=$10:01:21.0, $\delta_{\rm J2000}=$+55:53:50.5. Contour
  levels are from $5\sigma = 1.1$ to 4.7\,mJy\,beam$^{-1}$ in steps of
  0.6\,mJy\,beam$^{-1}$ for the 3.1\,mm continuum and from 3$\sigma =
  0.33$ to 0.77\,Jy\,\kms\ in steps of 0.11\,Jy\,\kms\ for the line
  emission. The synthesized beam of 3.2$''\times$3.0$''$ at
  p.a. $87^\circ$ is shown in the lower left corner (1$''$ corresponds
  to 9\,kpc at an angular-size distance $D_A = 1.87$\,Gpc). The
  white stars indicate the optical positions of the A and B
  image. The rms noise per 10\,MHz channel is $\sim $0.65\,mJy. The
  spectrum at the position of the jet is indicative of the noise
  level. Velocities are relative to the redshift at $z = 1.4141$.}
  \label{obs3mm}
\end{figure*}

\section{Observations}

CO(2-1) and (5-4) observations of Q0957+561 were made simultaneously
in 1998 and again in 2003 with the IRAM interferometer. The first
observation run is discussed in P99. We only give additional details
on the refined re\-duction of this data set. \\
{\sl Data taken in 1998:} The
observations were carried out in the B, C and D configurations. The
bandwidth of the cross-correlator was set up to cover 420\,MHz at both
frequencies with a spectral resolution of 2.5\,MHz, equivalent to
7.8\,\kms\ at 95.5\,GHz (redshifted CO(2-1)) and 3.1\,\kms\ at
238.7\,GHz (CO(5-4)). The bandpass was calibrated on 3C273, 3C345 or
0923+392 while phase and amplitude calibrations were performed on
0917+624. Unlike P99, we discarded data from an entire observing run
that were substantially reducing the quality of the maps. We have also
corrected for the velocity scale which was wrong by a factor of 1.5 in
P99 due to a data header error.\\
{\sl Data taken in 2003:} The observations were carried out in the C
and D configurations. The bandwidth of the cross-correlator was set up
to cover 580\,MHz with a spectral resolution of 1.25\,MHz
corresponding to 3.9\,\kms\ at 95.5\,GHz and 1.6\,\kms\ at
238.7\,GHz. We resampled the two data sets to a spectral resolution of
7.8\,\kms\ (2.5\,MHz) over the 420\,MHz band covered by the first
observations to ensure uniform noise in the merged velo\-city
channels. 3C273, 0851+202 and 0420-014 were used as bandpass
calibrators while 0804+499 and 1044+719 were taken as phase and
amplitude calibrators. Relative to P99, the sensitivity of the
combined data sets is now higher by a factor of $\sim$2.

%
%
\begin{table*}[htb]
\centering
 \begin{tabular}[c]{ccccccc}
 \hline
 \hline
  & \multicolumn {3}{c}{\hrulefill\hskip 3mm
    \raisebox{-.5mm}{CO(2$\rightarrow$1)}\hskip 3mm \hrulefill}
  & \multicolumn {3}{c}{\hrulefill\hskip 3mm
        \raisebox{-.5mm}{Continuum at 3.1\,mm$^b$} \hskip 3mm \hrulefill}\\
  &  A$-$blue & A$-$red & B$-$blue & A & B & C \\
 \hline
 Rel. positions ($''$) & ($-$1.9,$+$3.8) & ($-$1.6,$+$6.1) &
 ($+$0.2,$-$1.1) & ($-$1.45,$+$5.00) & ($-$0.24,$-$1.0) & (3.23,7.13)\\
 Position errors ($''$)$^a$ & (+0.2,+0.2) & (+0.3,+0.3) & (+0.2,+0.2) &
 (+0.06,+0.08) & (+0.08,+0.10)& (+0.07,+0.08) \\
 Flux density (mJy) & $-$ & $-$ & $-$ & $4.7\pm0.2$ & $3.4\pm0.2$ & $6.2\pm0.2$ \\
 Peak flux$^c$ (mJy/beam) & 0.9$\pm$0.2 & 2.1$\pm$0.2  & 2.2$\pm$0.2
 &  4.6$\pm$0.2  &  3.3$\pm$0.2  &  5.1$\pm$0.2 \\
 & (1.1$\pm$0.3) & (1.7$\pm$0.3) & (2.3$\pm$0.3) & (4.2$\pm$0.3) & (2.5$\pm$0.3) & (5.5$\pm$0.3) \\
 FWHM  (\kms)  & 280$\pm$60 & 160$\pm$20  & 280$\pm$50  &  $-$ & $-$ & $-$ \\
 Vel.\ at line peak$^d$ (\kms) & $-$140$\pm$30 & 340$\pm$15  & $-$140$\pm$23  & $-$ & $-$
 & $-$ \\
 Integr.\ line flux (Jy\,\kms) & 0.25$\pm$0.06 & 0.34$\pm$0.06 & 0.61$\pm$0.06
   & $-$ & $-$ & $-$ \\
 \hline

  \end{tabular}
  \caption{Flux densities, velocities and positions of individual
  image components. Values have been determined by fitting Gaussian
  profiles to the visibilities. Offset position (0,0) is at the
  assumed lens position specified in the caption of
  Fig.\ref{obs3mm}. $^a$ Relative position uncertainties are based on
  statistical noise and do not include astrometric errors. $^b$ Values
  taken from the second observation campaign in 2003. $^c$ Values in
  brackets are from the reanalysis of the P99 data, and do not include
  data from 2003. $^d$ Velocities are relative to the redshift at $z =
  1.4141$. The CO(2--1) line parameters for the two A components were
  not derived from the spectrum shown in Fig.\ref{fig2} that was taken
  {\it between} A-red and A-blue, but are based on spectra determined
  at the respective positions of A-blue and A-red.}
\label{obsval}
\end{table*}
%

\section{The data}

\subsection{Continuum emission}
{\sl 3.1\,mm:}\,\,\,\, The radio continuum was computed separately for
the two data sets by averaging the line-free channels with a
velo \-city lower than $-510$\,\kms\ and higher than $+540$\,\kms. Two
lensed images of the quasar, labeled A and B, and a radio jet C appear
at this wavelength (Fig.\ref{obs3mm}), in agreement with VLA
observations (Harvanek et al.\ \cite{harv}). The positions coincide
with the optical ones within the errors (Tab.\ 1).  Also in agreement
with previous work, the A and B components are pointlike whereas the
radio jet C component is slightly elongated ($\sim 2.7''$, PA
$\sim$60$\grad$) in the beam of the interferometer.  Finally, there is
no evidence for variability above 10\% between May 1998 and April 2003
in all three components. This coincides with results obtained at 3.6cm
with the VLA by Harvanek et al.\ (\cite{harv}).\vskip 1.1mm

\noindent {\sl 1.3\,mm:}\,\,\,\, A noise level of $\sigma=0.6$\,mJy
was obtained in the 2003 data by averaging over the entire
bandwidth available at 1.3mm. We find a 3.5$\sigma$ peak at the 3mm
position of the A-continuum component, a 2$\sigma$ peak at the
position of the B component and a $4\sigma$ peak at the position of
C. The steep drop in flux density at 1.3 mm is consistent with the
weakness of the synchrotron emission expected if we extrapolate to
this short wavelength the spectral index of $\simeq\;-0.6$ measured in
the 20 mm (Harvanek et al 1997) to 3 mm (P99, and this paper) range.
Any dust emission is either too weak or too extended (say $>3''$) to
be detected at this frequency. Data from the 1998 run were not taken
into consideration as they were much less sensitive.

\subsection{Line emission}
{\sl CO(2-1):}\,\,\,\, To estimate flux densities in the CO(2-1) line
(Tab.\ 1), the 3.1\,mm continuum was first removed separately from
each data set to account for possible low level continuum
variability. Then, all data sets were combined to a single uv-table,
and velocity channels in the $-510$\,\kms\ to $+540$\,\kms\ range
was summed to produce an integrated line map (Fig.\ref{obs3mm}). The
resulting map shows two lensed images labeled CO$-$A and CO$-$B
(P99). The emission centroids of the two images are separated by
$\simeq 7''$, about $1''$ more than in the optical, radio and
millimeter continuum (Tab.\ 1). The spectral profiles taken towards
positions CO-A and CO-B are different. While a double-peaked profile
is visible towards the northern image CO-A, a single blueshifted
velocity component is detected towards CO-B. The shape of the blue
components measured towards CO-A and CO-B is similar. The CO$-$F image
reported by P99 could not be confirmed with the new
observations.\vskip 1.5mm

\noindent {\sl CO(5-4):}\,\,\,\, No significant line emission was
detected at the position of CO$-$A and CO$-$B, the two line components
detected at 3mm. A $\sim 1.9$\,mJy ($=3\sigma$) peak is only
tentatively detected in the velocity integrated map ($-300$\,\kms\ to
$+100$\,\kms) at the position of CO$-$B. The intensity of the CO$-$B
component at 3mm over the same velocity range is estimated to be
1.8\,mJy. Although uncertain because of the low signal to noise ratio
of the (5-4) line, and possibly because of some residual continuum
emission, we set an upper limit of $\leq 1$ on the velocity averaged
intensity ratio $R_{54}=$CO(5-4)/CO(2-1) toward CO$-$B, the strongest
of the two line components.

\begin{table*}
\centering
\footnotesize
\begin{tabular}{ccccccc}
\hline
\hline
     & SPLS $^e$ & SPEMD $^e$ & SPEMD+CL $^f$ & FGS $^e$ & FGSE $^e$
 & FGSE+CL $^e$ \\
\hline
Source Positions \\
 (Source plane) \\
\multicolumn{1}{c}{{\it Radio continuum}} \\
$(\Delta\alpha_s,\Delta\delta_s)$ ($''$) $^a$  & ($-$0.02,0.90) &
 ($-$0.14,1.02) & (2.20,2.49) & ($-$0.15,1.06) & ($-$0.18,1.10)
 & (1.80,2.58) \\
\multicolumn{1}{c}{{\it Line emission}}\\
$(\Delta\alpha_s,\Delta\delta_s)$ ($''$) $^a$  & ($-$0.23,0.95)  &
 ($-$0.32,1.00) & (1.93,2.39) & ($-$0.38,1.05) & ($-$0.40,1.05) &
 (1.54,2.47) \\
\hline
Lens properties\\
\multicolumn{1}{c}{{\it Galaxy:}} \\
$\sigma_v$ (\kms) & 400 & 400 & 360 &300 & 340 & 320 \\
$\xi_c$ ($''$) & 0.12 & 0.12 & 0.02 & 1.2 & 0.6 & 1.1 \\
$\varepsilon$ & $-$ & 0.1 & 0.2 & $-$ & 0.1 & 0.3 \\
$\Phi_{\rm PA}$ ($\grad$) $^b$ & $-$ & {\it 64} & {\it 64} & $-$ &
	    {\it 64} & {\it 64} \\
$\eta$ & 1.1 & 1.0 & 1.1 & $-$ & $-$ & $-$ \\
\multicolumn{1}{c}{{\it Cluster}:}\\
$\gamma'$ $^{d}$ & 0.3 & 0.2 & $-$ & 0.4 & 0.3 & $-$ \\
$\Phi_{\gamma'}$ ($\grad)$ $^b$  & {\it 60} & {\it 60} & $-$ & {\it
  60} & {\it 60} & $-$ \\
$\sigma_{\rm cl}$ (\kms) & $-$ & $-$ & 375 & $-$ & $-$ & 350 \\
$(\Delta\alpha_{\rm cl},\Delta\delta_{\rm cl})$ ($''$) $^b$ & $-$ & $-$ &
 {\it (13.7,6.9)} & $-$ & $-$ & {\it (13.7,6.9)} \\
\\
\multicolumn{1}{c}{{\it Black hole}:}\\
$M_{\rm bh}$ ($10^{12}\Msun$) & $-$ & $-$ & $-$ & 0.5 & 0.3 & 0.2 \\
\hline
$\Delta\tau_{\rm cont}$ [days] & 411 & 420 & 417 & 417 & 421 & 416 \\
$\Delta\tau_{\rm line}$ [days] $^g$ & 421 & 450 & 412 & 423 & 454 & 414 \\
$m_{\rm RC}^{\rm total}$ &  5  &  4 &  4 &  3 &  3 &  3 \\
$m_{\rm CO}^{\rm total}$ &  6  &  7 &  9 &  5 &  6 &  6 \\
\hline
$N_{\rm dof}$ $^{c}$  & 3 & 2 & 2 & 3 & 2 & 2 \\ 
$\chi^2_{\rm r}$ $^{c}$ & 3.0 (3.4) & 5.5 (40) & 1.5 (1.5) & 1.3
(1.2) & 5 (40) & 2.0 (2.5) \\
\hline
\end{tabular}
\caption{Best fit parameters for Q0975+561. Parameters are:
$\Delta\alpha\equiv$ right ascension, $\Delta\delta\equiv$
declination, $\sigma\equiv$ velocity dispersion, $\xi_c\equiv$ core
radius, $\varepsilon\equiv$ ellipticity, $\Phi_{\rm PA}\equiv$
position angle, $\eta\equiv$ power law index, $\gamma^{(_{'})}\equiv$
shear term, $\Phi_\gamma\equiv$ shear angle, $M_{\rm bh}\equiv$ mass
of the black hole, $\Delta\tau\equiv$ time delay, $m\equiv$
magnification factor.  $^a$ Offset from assumed lens position. $^b$
Fixed parameters (shown in {\it italic}). $^c$ Values in brackets are
$\chi_r^2$-tests with time delay. The reduced $\chi^2_{\rm r}$ is
defined as: $\chi^2_{\rm r}=\chi^2/N_{\rm dof}$ where $N_{\rm dof}$ is
the degrees of freedom. $^d$ $\gamma'$ is only taken for the
elliptical potentials ($\gamma'=\gamma/(1-\kappa)$) where
$\kappa=\Sigma_{\rm clus}/\Sigma_{\rm crit}$ is defined as the surface
mass density of the cluster in units of the critical surface density
(Barkana et al.\ 1999). For the spherical models, $\kappa=0$ and for
the elliptical ones it can be approximated by $\kappa\simeq \gamma$,
i.e. $\gamma'=\gamma/(1-\gamma)$.  $^e$ based on Barkana et
al. (1999). $^f$ based on Chae et al.\ (1999). $^g$ determined for the
blue component.}
\label{fitpar}
\end{table*}

\section{Modelling Q0957+561}
We developed a numerical code based on the standard gravitational
lens equation (Schneider et al.\ \cite{schn}) to explain the
absence of the double-peaked line profile toward CO$-$B and to
investigate the distribution and kinematics of the molecular gas
around Q0957+561.

The code has been applied on 6 different models of Q0957+561 (Tab.\
2), all based on previous work by Barkana et al.\ (\cite{bark}) and
Chae et al.\ (\cite{chae}). The mass distributions of three models are
based on King profiles, as proposed by Falco et al.\ (\cite{falc}$-$
models: {\footnotesize FGS, FGSE, FGSE+CL}), the remaining three on a
softened power-law distribution, as suggested by Grogin et al.\
(\cite{grog}$-$models: {\footnotesize SPLS, SPEMD, SPEMD+CL}).
{\footnotesize FGS} and {\footnotesize SPLS} are spherical models
where the effects of the surrounding cluster at $z=0.355$ are
approximated by an external shear. {\footnotesize FGSE} and
{\footnotesize SPEMD} models take into account the ellipticity in the
lens galaxy, the remaining models use a single isothermal sphere
({\footnotesize SIS}) to model the lens properties of the cluster
({\footnotesize SPEMD+CL, FGSE+CL}). A point mass was added to the
King profiles to account for a black hole in the center of the lens
galaxy (Mediavilla et al.\ \cite{medi}; Barkana et al.\ \cite{bark}),
except for the softened power law models which implicitely cover this
case. The group of background galaxies at $z= 0.5$ has not been
considered, however. The composite pseudo-Jaffe models by Keeton et
al.\ (2000) have not been taken into account as they would not have
provided further details for our analysis.

To find the set of parameters for each model reproducing our
observations with the lowest $\chi_r^2$, we first restricted the
parameter space with simple assumptions based on the number of lensed
images, separations among them etc. We then scanned the parameter
space for each model to get the lowest $\chi_r^2$ and included the
following constraints: relative positions of the A and B
continuum components at 3.1mm with respect to the lens (4
constraints), relative positions of CO-A-blue and CO-B-blue
(Tab.\ref{obsval}), the continuum and line intensity ratios and the
optical/radio time delay of 400-420 days from Kundic et al.\ (1997)
and Haarsma et al.\ (1999). This results in a total of 11
constraints. For simplicity, compact fixed-size Gaussians were used
to approximate the respective components. The so found best-fit models
turned out to be in excellent agreement with simulations based on VLA
and optical data (e.g. Barkana et al.\ \cite{bark}, Chae et al.\
\cite{chae} and Keeton et al.\ 2000).

The {\footnotesize SPEMD+CL}, {\footnotesize FGSE+CL} and the
{\footnotesize FGS} models reproduce the observed constraints with the
lowest $\chi_r^2$ ($\leq3$; Tab.\ \ref{fitpar}). Although the
{\footnotesize FGS} model yields one of the best results, we do not
favour spherical mass distributions because of the ellipticity of the
lens (Bernstein et al.\ \cite{bern}).  The contribution of the cluster
is also important: The models that best explain these observations all
require an {\footnotesize SIS} cluster ({\footnotesize FGSE+CL} and
{\footnotesize SPEMD+CL}, the latter has a lower
$\chi^2$)\footnote{For the following discussion, we will mainly use
the best-fit {\footnotesize FGSE+CL} model since it provides the
closest match to already published results, e.g.\ Keeton et al.\
(2000).  However, both models are fully equivalent for argumentation
purposes.}. We note that the time delay derived from the same models
are in rather good agreement with recent optical and radio
measurements ($\sim 400-420$ days; Kundic et al.\ 1997; Haarsma et
al.\ 1999).

\subsection{Line emission}
In the following two subsections we discuss different approaches for
modelling the blueshifted line emission based on the best-fit
{\footnotesize FGSE+CL} potential and then explain the difference seen
in the profiles toward CO$-$A and CO$-$B.

\begin{figure}[t]
     \centering
     \resizebox{\hsize}{!}{\rotatebox{-90}{\includegraphics{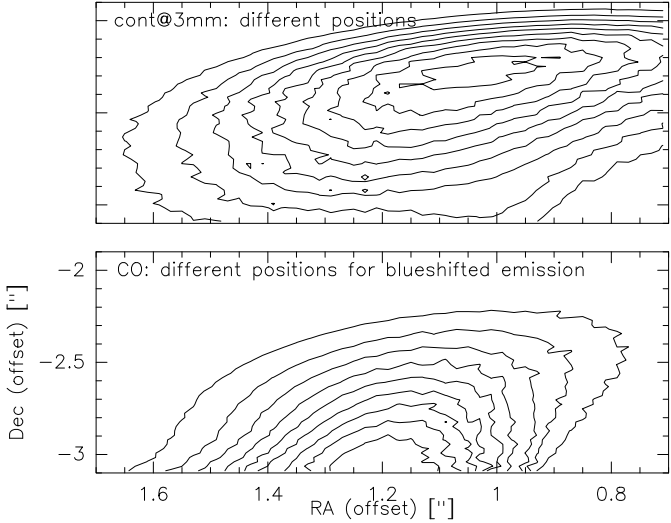}}}
     \caption{$\chi^2$ results for the continuum ({\it upper panel})
     and blueshifted line ({\it lower panel}) position fits based on
     the {\footnotesize FGSE+CL} model. Shown here are different
     positions in the source plane for the blueshifted line emission
     and the continuum depending on the $\chi^2$-test including the
     observed intensity ratios between the two lensed images and the
     observed position.  Assumed values for the 3mm continuum:
     A/B$=1.5$; blueshifted line: A/B$=0.4$. Positions for the A, B
     (continuum) and CO$-$A and CO$-$B can be found in
     Tab.\ref{obsval}. Contour levels are from 0.1 to 1.0 in steps of
     0.1 (in terms of exp(-$\chi^2$)). The region with a value close
     to 1.0 in this map corresponds to the best-fit position.}
     \label{posAB}
\end{figure}

\begin{figure}[!]
     \centering
     \resizebox{8cm}{!}{\rotatebox{-90}{\includegraphics{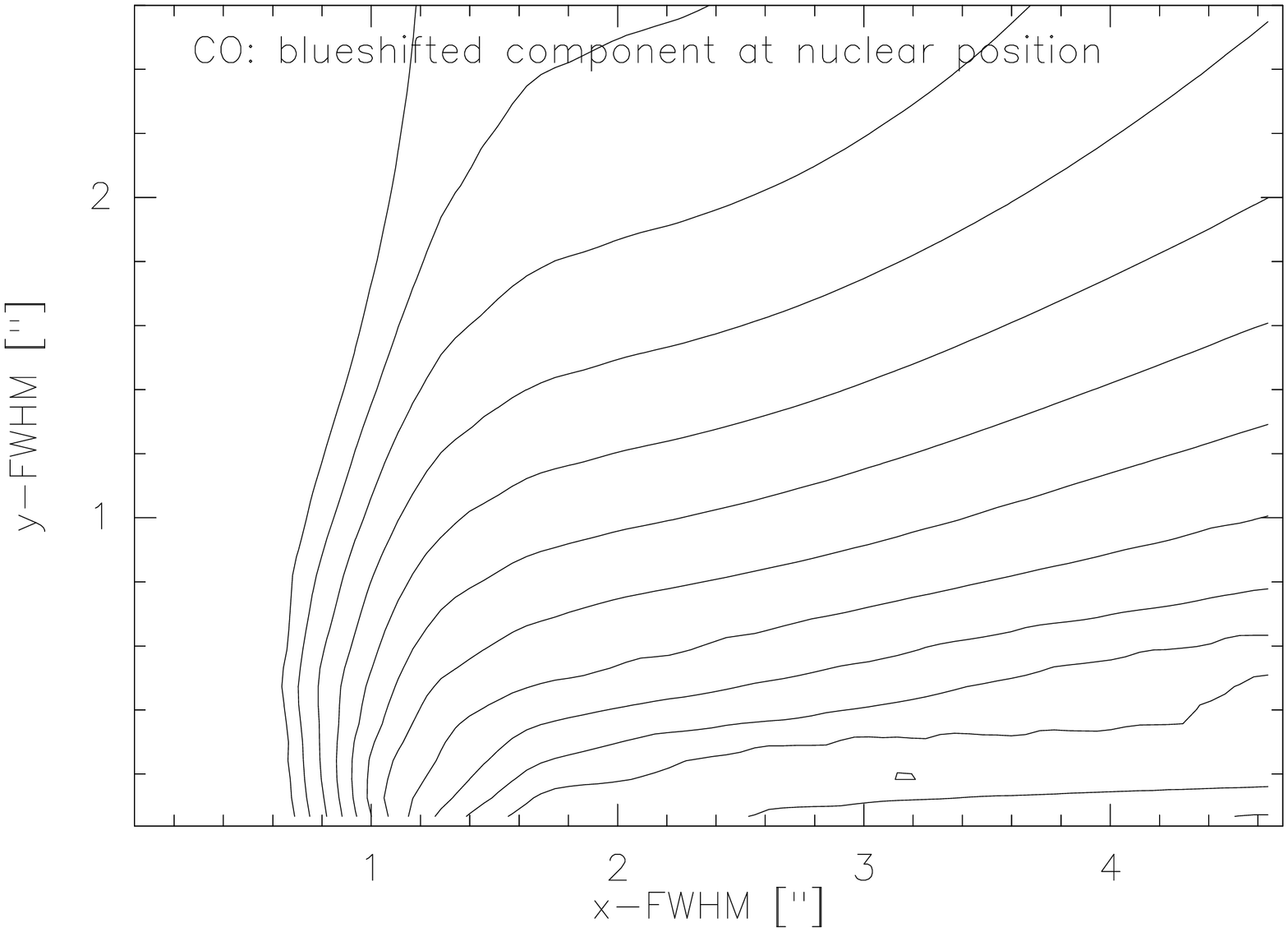}}}
     \resizebox{8cm}{!}{\rotatebox{-90}{\includegraphics{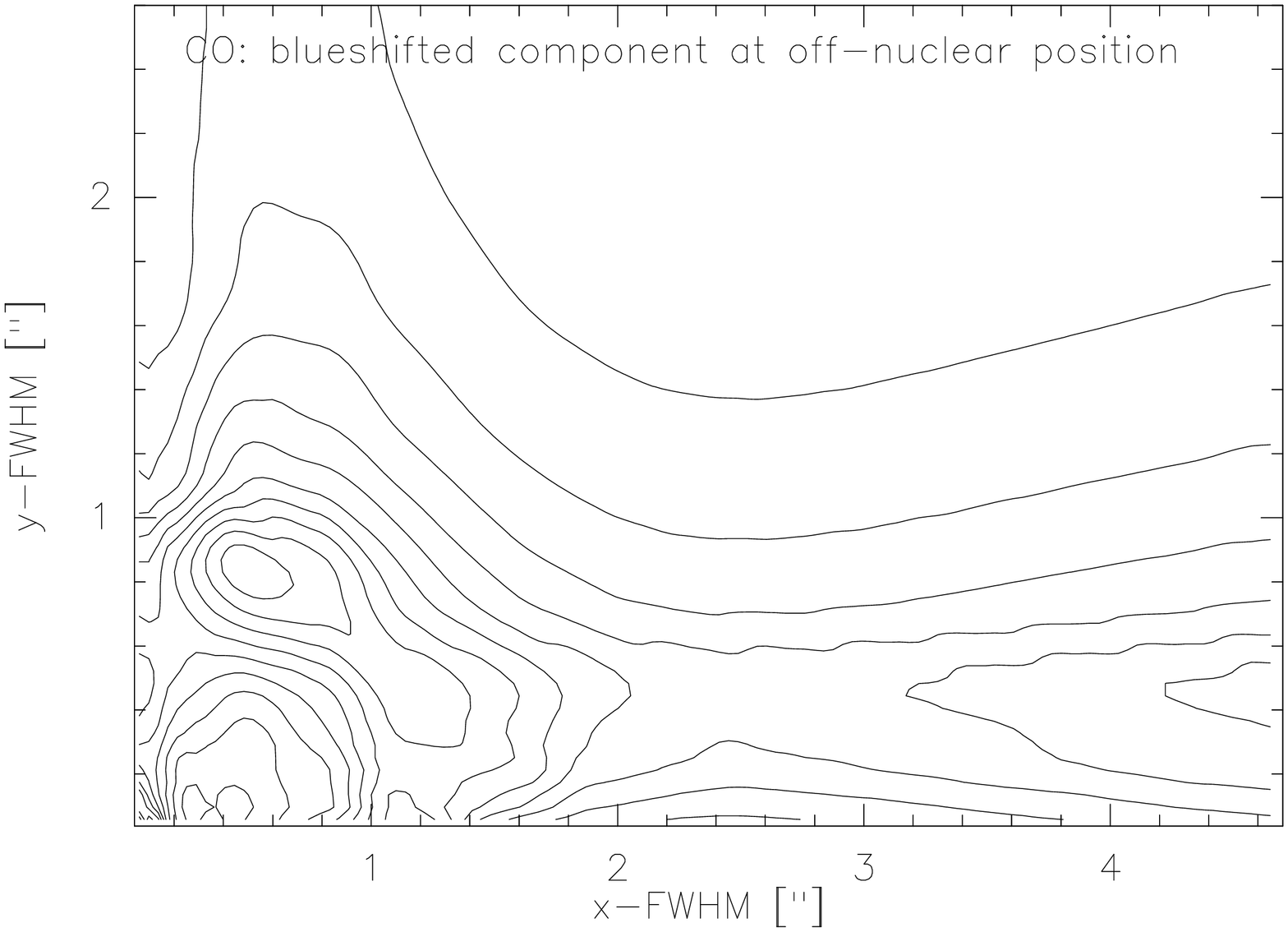}}}
     \resizebox{8cm}{!}{\rotatebox{-90}{\includegraphics{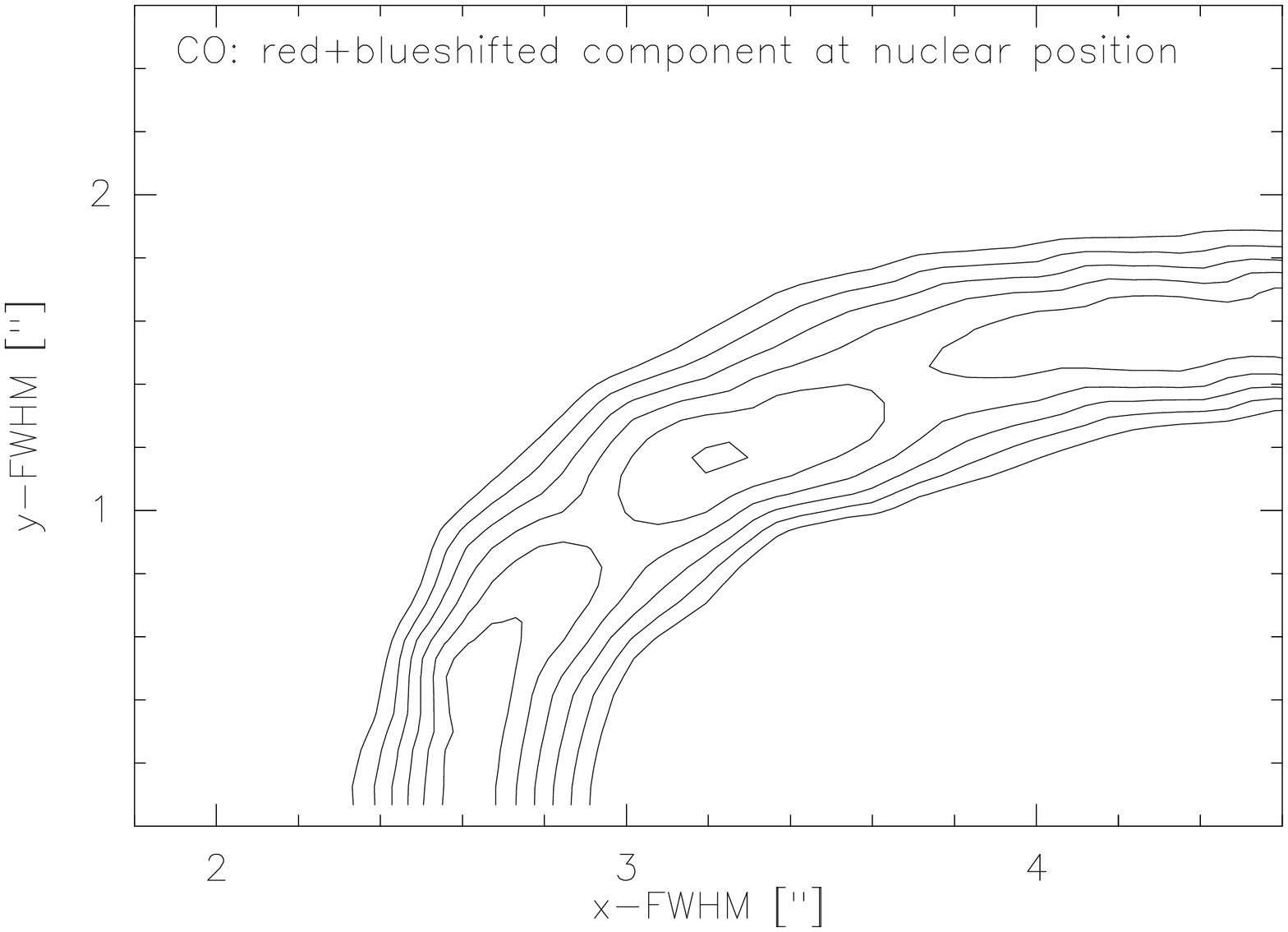}}}
     \caption{$\chi^2$ results of the size fit for the line
     emission. The FWHM of the Gaussian profile are plotted with the
     respective $\chi^2$. We also varied the position angle (PA) from
     5 to 60$\grad$ but we restricted the presentation to the PA with
     the lowest $\chi^2$. {\it Upper panel:} Blueshifted emission
     centered on the nucleus. $\chi^2$ values based on the
     A-blue/B-blue ratio. Lowest $\chi^2$ for PA=20$\grad$. Contour
     levels start at 0.56 to 1.0 in steps of 0.04. {\it Middle panel:}
     Blueshifted emission centered at a fixed off-nuclear position
     closer to the tangential caustic ($ \equiv$position of line
     emission in Tab.\ref{fitpar}). $\chi^2$ based on the
     A-blue/B-blue ratio. Lowest $\chi^2$ for PA=20$\grad$. Contour
     levels are from 0.1 to 1.0 by 0.1 (in terms of
     exp(-$\chi^2$)). {\it Lower panel:} Red- and blueshifted emission
     centered on the nucleus. $\chi^2$ values based on the
     (A-blue+A-red)/B-blue ratio. PAs of 50$\pm10\grad$ all result in
     a low $\chi^2$. Contour levels are from 0.6 to 1.0 in steps of
     0.08.}
     \label{size}
\end{figure}

\subsubsection{The blueshifted line}
\label{secblue}

Let us first concentrate on the blueshifted line. We can propose two
different scenarios for the origin of blueshifted line emission:
\begin{itemize}
\item a compact or extended\footnote{Simulated by an elliptical
  Gaussian profile, i.e.\ a distribution which is symmetric with
  respect to its centroid.} region centered on the nucleus,
\item a compact or extended region spatially not coincident with the
nucleus.
\end{itemize}
The small line ratio A-blue/B-blue of $\sim 0.4$ already rules out a
compact region centered on the nucleus, since it disagrees with the
measured 3mm continuum ratio A/B of $1.5\pm 0.1$. Even if we account
for a contribution of the jet in the A component and assume a lower
ratio for the core emission at 3mm, we know from VLBI observations
that the ratio A/B at the core is close to ~1.3 (Garret et al.\
1994). Thus, the continuum at 3mm and the blueshifted CO emission
cannot originate from the same compact region (compare also
Fig.\ref{posAB}).

On the other hand, an extended line emitting region centered on the
nucleus can indeed result in an A-blue/B-blue of 0.4. To constrain the
size of the blueshifted region, we assumed an elliptical Gaussian
profile and fitted the full widths at half maximum (FWHM) and the
position angle (PA) depending on the continuum and line ratios between
the lensed images. A best-fit (Fig.\ref{size}) is obtained for a
disk-like region, that is extended in the direction of the nuclear jet
($\sim 5''$) and tiny in the direction perpendicular to the jet
($\sim 0.2''$, PA=20$\grad$), with the south-western part of the disk
crossing the inner tangential caustic of the lens. A priori, such a
model produces multiple images (up to 5; compare also Keeton et al.\
2000) that result in a stronger CO$-$B image.

Alternatively to the extended region centered on the nucleus,
blueshifted emission from a rather {\it compact} region ($<1''$) close
to the inner tangential caustic and slightly off ($\sim0.5''$ to the
south-west) the nuclear position is also consistent with a line ratio
of $\sim 0.4$ (Fig.\ref{size}). A symmetric {\it extended} region
($>1''$) at an off-nuclear position does not produce such a low
A-blue/B-blue of 0.4.

Thus, an independent blueshifted emission must either originate
in an extended region centered on the nucleus or in a compact region
with a position close to the tangential caustic.

\subsubsection{The redshifted line}

\begin{figure*}[!]
     \centering \vspace*{0cm}
     \resizebox{\hsize}{!}{\rotatebox{-90}{\includegraphics{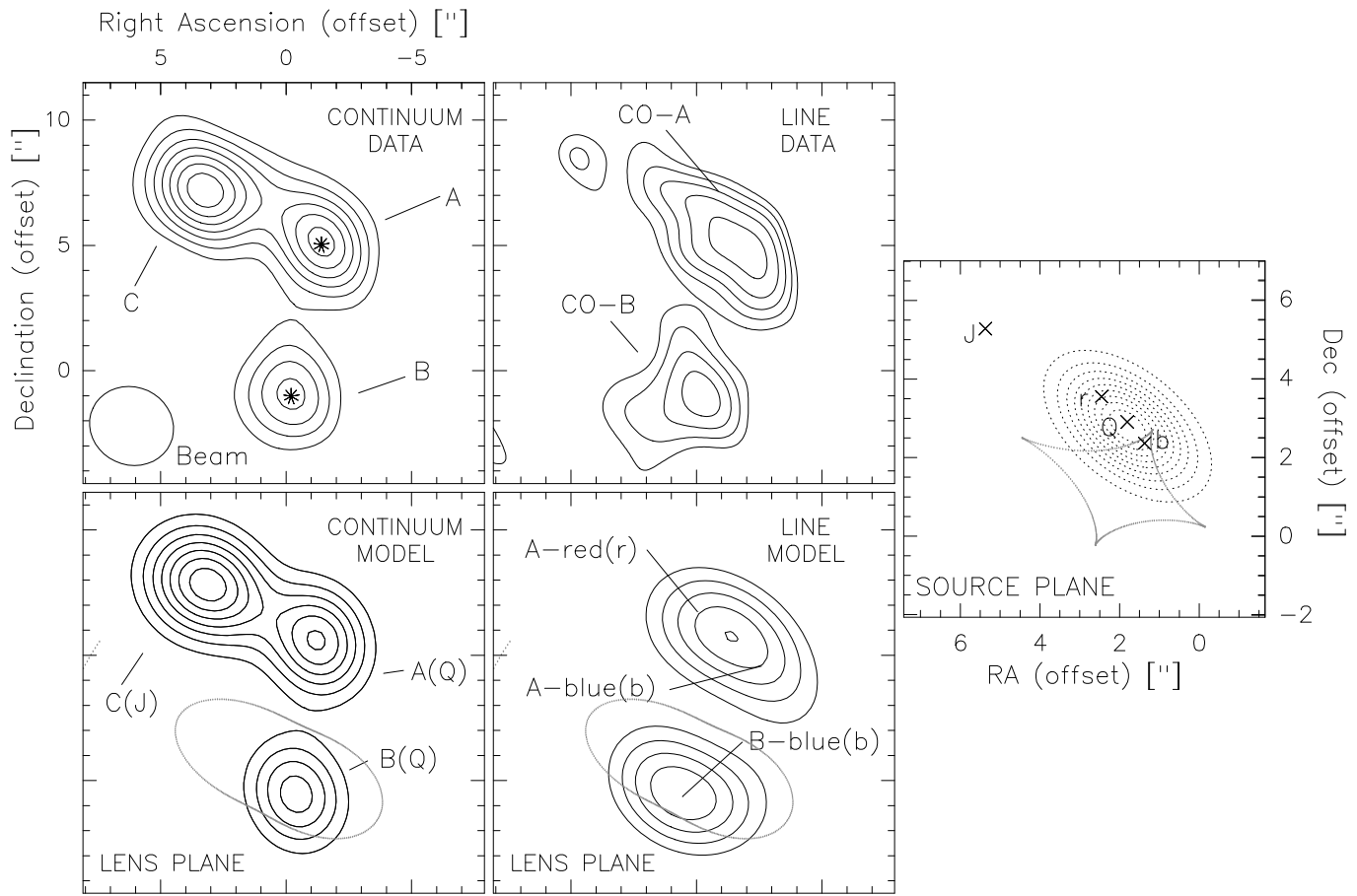}}}
     \caption{The line-free continuum at 95\,GHz ({\em upper left
     panel}) and the $^{12}$CO(2$\rightarrow$1) line emission ({\em
     upper middle panel}) as observed with the IRAM interferometer
     from both data sets. The lower panels show the {\it simulated}
     radio continuum ({\em left}) and line emission ({\em middle})
     convolved with the synthesized beam  and the critical lines ({\it
     grey lines}). In the {\it right} panel the location of the {\it
     unlensed} line emissions (r$\,\equiv \,$A$-$red,b$\,\equiv\,
     $B$-$blue$\, \equiv\,$A$-$blue) as well as the {\it unlensed}
     radio continuum (J$\,\equiv\,$jet, Q$\,\equiv\,$quasar) is shown
     together with the radial and tangential caustics with respect to
     the assumed position of the lensing galaxy ({\it grey
     lines}). The offset position (0,0) of each map corresponds to the
     position of the lensing galaxy. Contour levels for the data are
     as Fig.\ref{obs3mm}. We have also used the same contour levels
     for the simulated maps. The black stars indicate the optical
     positions of the A and B image. The parameters used for the model
     ({\footnotesize FGSE+CL}) are listed in Tab.\
     \ref{fitpar}. Changes in the velocity dispersion of the lens of
     $\sim$ 1\% or in the source positions of $\sim 0.1''$ produce a
     difference in the time delay of the {\footnotesize FGSE+CL} model
     of $\sim$ 5\%.}
     \label{fig2}
\end{figure*}

The difference in the line profiles observed toward CO$-$A and CO$-$B
can be explained by the location of the redshifted gas component
relative to the lens caustic. The best-fit simulation for the
{\footnotesize FGSE+CL} model produces one tangential caustic
(Fig.\ref{fig2}). The radial caustic is at infinity. While the
continuum and blueshifted line components remain close to the
tangential caustic and are thus deflected into two distinct images,
the redshifted component is already located too far off the caustic
(to the north-east) to generate two lensed images. Thus, it appears
that both blue- and redshifted gas become visible toward CO$-$A while
blueshifted gas only can be detected toward CO$-$B. The unlensed
separation between the centroids of the blue- and the redshifted
components, corrected for magnification, are estimated to $\sim$1$''$,
or equivalently $\sim 9$\,kpc at the distance of Q0957+561.

Likewise, an extended region centered on the nucleus where the
north-eastern part corresponds to the redshifted component and the
south-western part to the blueshifted one produces equivalent results
(Fig.\ref{size}). To obtain the derived line ratio
(A-blue+A-red)/B-Blue $\sim 1.0$, the size of the region has to be
extended by about $\sim 3-4''$. This value seems to be independent of
the lens potential even if we find that the SPEMD+CL model appears to
favour slightly more elliptical profiles for the emission than the
FGSE+CL model (cf. also Keeton et al.\ 2000). If we divide the lensed
CO$-$A into a red- and blueshifted part and derive then the ratio
between A-blue and B-blue, we again obtain a value of $\sim
0.4$\footnote{In contrast to Section\,\ref{secblue}, where we found
that an extended blueshifted emission region would not result in an
A-blue/B-blue ratio of 0.4 if symmetrically centered on a position
closer to the tangential caustic, the extended emission region for the
blueshifted part is not symmetric with respect to the position used
there. This might explain why we nevertheless obtain such a low ratio
for the blueshifted part alone.}.

The sensitivity is not sufficiently high to establish the absolute
genuineness of the CO$-$A-blue component from the observations alone,
but our best-fit {\footnotesize FGSE+CL} model of the host galaxy not
only agrees with the Barkana model of Keeton et al.\ (2000), but a
priori corroborates the detection of blueshifted CO gas in the
direction of component A.

\subsection{Radio continuum}
The morphology and the intensity ratios of the radio continuum can
easily be modelled. As already noted by Schneider et al.\
(\cite{schn}), the radio jet J is lensed only once due to its location
far off the caustic.

\begin{figure}[t]
     \centering
     \rotatebox{-90}{\resizebox{\hsize}{!}{\includegraphics{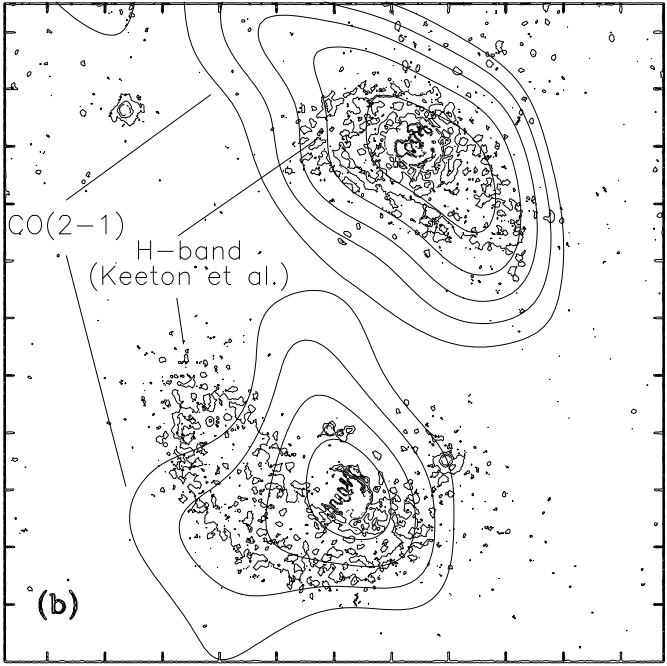}}}
     \caption{HST image in the H-band of the host galaxy in Q0957+561
            (Keeton et al.\ 2000) with contour levels from Fig.\ref{fig2}
            of the velocity integrated CO(2$-$1) emission.  Separation
            of the tick marks is 1$''$.}
     \label{keet-plan}
\end{figure}

\section{Discussion}
The 3mm continuum observations of the redshifted double quasar
Q0957+561 are in good agreement with observations at other
wavelengths.  While the 3mm continuum emission from the quasar is
unresolved, emission from the radio lobes is extended along the
direction of the jet. The CO(2-1) line, however, does not originate in
the quasar. The difference in position, and therefore in the
deflection, explains also the difference between the flux density
ratios in the 3mm continuum (A/B $=1.5$ i.e.\ like in the radio
continuum) and in the line (CO$-$A/CO$-$B $=0.4$ for the blue
component), and accounts for different time delays for the CO(2-1) and
the radio images. We note, however, that the time delay difference is
only marginally significant (at most 5$-$20 light days, according to
the models).

What can we conclude about the origin of the redshifted CO(2-1)
velocity component? \hskip 3mm P99 argued that it was either
originating from a companion galaxy, very close to Q0957+561 (see also
Papadopoulos et al.\ \cite{papa}) or tracing the presence of molecular
gas in the rotating disk of the host galaxy.  Our models discard the
first hypothesis.  First, there has been no detection of such a
companion galaxy at other wavelengths, but the host galaxy, which has
already been revealed in HST observations by Keeton et
al. (\cite{keet}), is oriented in the direction of the two line
components in CO$-$A and appears to be extended by $\sim 3''$
(Fig. \ref{keet-plan}). Furthermore, models for the gas emission in
the host galaxy that take the velocity profiles observed toward CO$-$A
and CO$-$B into account are similar in size and shape (see
Fig.\ref{fig2}) to those invoked by Keeton et al.\ (2000). The
blue- and redshifted CO(2--1) emission appear to be very likely
connected to the stellar distribution of the host galaxy and thus
corroborate the results of Keeton et al.\ (2000), whereas models based
on a single blueshifted component favour a very thin, elongated disk
or a compact CO(2--1) emitting region slightly off the nuclear region,
and as such are rather in contrast with images of the host galaxy.
Moreover, the integrated line ratio (CO$-$A-red+CO$-$A-blue)/CO$-$B
$\sim 1.0$ is consistent with the ratio H-band-A/H-band-B$\sim$1
derived by Keeton et al.\ (\cite{keet}). This is what one would expect
if the profile of the gas distribution is very similar to the stellar
one. Also, the double-peaked line profile at CO$-$A which appears to
be symmetric and centered within the errors at $z=1.4141$ is a
classical signature for rotation and therefore, provides further
support that the CO emission is associated with the host galaxy.  By
the same arguments we are tempted to exclude also a recent galaxy
merger in Q0957+561, though this possibility cannot be ruled out
completely.\\

Based on these arguments, we favour P99's second hypothesis: the
presence of an important reservoir of molecular gas in the disk of the
host galaxy surrounding Q0957+561.

According to the models the continuum is less magnified by a factor of
2$-$3 relative to the strongly lensed optical images ($\sim 10$) while
the total magnification factor derived in the CO line varies between
$m_{\rm\small CO}^{\rm total}\simeq 6$ ({\footnotesize FGSE+CL}) and
$m_{\rm CO}^{\rm total} \simeq8$ ({\footnotesize SPEMD+CL}). Based on
the lower magnification, we have estimated an upper limit to the
molecular gas mass for the integrated blue- and redshifted velocity
profile in CO$-$A and CO$-$B. The {\footnotesize FGSE+CL} model
corresponds to the higher CO$-$luminosity case and provides therefore
a slightly higher molecular gas mass. Under the assumption that the
brightness temperature of the (1-0) and (2-1) lines is the same, we
adopt a CO to H$_2$ conversion factor of $\simeq 4.8\,{\rm M}_\odot$
(K\,km\,s$^{-1}$pc$ ^2)^{-1}$ based on a determination for the Milky
Way (Solomon \& Barrett \cite{solo91}). Clearly, the Milky Way and the
Q0957+561 host galaxy are in different evolutionary stages and show
different properties, so the adopted conversion factor is likely to be
an overestimate (cf. Downes et al.\ 1993), and an upper limit for the
molecular gas mass will be obtained. Based on the FGSE+CL
magnification factor, we estimate M$_{\rm gas}$=M(H$_2$+He)$\simeq
2.4\times10^{10}\,{\rm M}_\odot$ for the blue profile and $\sim
2.7\times10^{10}\,{\rm M}_\odot$ for the reddened profile ($\simeq
9\times10^9\,{\rm M}_\odot$ for {\footnotesize SPEMD+CL}), giving an
upper limit for the total gas mass of $\simeq 5.1\times10^{10}\,{\rm
M}_\odot$. We can also give a lower limit for the gas mass of M$_{\rm
gas}\sim 8\times 10^9$M$_\odot$ assuming that the CO(1--0) line is
optically thin (Solomon et al.\ \cite{solo97}). This relatively small
difference of only a factor of 6 indicates that the standard
conversion factor might nevertheless already give a reliable estimate
of the gas mass in the host galaxy around Q0957+561. Assuming a radius
of $\sim$10kpc and a velocity separation between both line profiles of
$\sim 400$\kms, we find a dynamical mass of $\sim 4\times10^{11}{\rm
M}_\odot$ (not including inclination effects). This is about ten times
higher than the derived gas mass consistent with what is found for
other active galaxies (e.g. Evans et al.\ 2002, Sakamoto et al.\
1999). The low upper limit on the velocity-averaged line intensity
ratio, $R_{54} \simeq 1$, favours the hypothesis of low excitation CO
emitted mainly from the disk of the host galaxy.  The agreement
between molecular gas masses obtained with the integrated CO
luminosities and individual magnification factors, each tracing line
emission from half of the quasar host, is further support for the
rotating disk hypothesis.  If our assumptions are correct, the
mole\-cular gas in the disk cannot be very hot, but to our know\-ledge
no sensitive observations in the ground and higher rotational CO
transitions have been carried out yet to confirm this conclusion.

\begin{table}
\centering
\hspace*{-0.1cm}
\begin{tabular}{cccc}
\hline
\hline
   & CO$-$A-red & CO$-$A-blue & CO$-$B-blue \\
\hline
$mL'_{\rm CO}$ $^a$&&& \\
{\tiny(10$^{10}$K\kms pc$^2$)} 
& 0.8$\pm$0.3 & 0.8$\pm$0.2 & 1.8$\pm$0.2\\[0.1cm]
$m_{\rm FGSE+CL}$ $^b$       & 1.5 & 1.7 & 4.3 \\[0.1cm]
$M_{\rm gas}$ (10$^{10}$ M$_\odot$)& 2.7$\pm$1.0 & 2.4$\pm$0.6 & 2.3$\pm$0.6 \\
\hline
 total gas mass & \multicolumn{3}{c}{\raisebox{-0.5mm}{$M_{\rm gas}^{\rm tot}
 \equiv M_{\rm gas}^{\rm red}+M_{\rm gas}^{\rm blue} \simeq 5\times 10^{10}$
  M$_\odot$ }}  \\[0.1cm]

\hline

\end{tabular}
\caption{Apparent CO luminosity $L'$ and molecular gas mass $M_{\rm
gas}$ of Q0957+561.  $^a\,m\equiv$ magnification
factor. $^b$\,determined with the {\footnotesize FGSE+CL} model. }
\end{table}


\section{Conclusions}

Recent sensitive observations were combined with earlier data by
Planesas et al (1999) to corroborate their original discovery of
CO(2-1) line emission from Q0957+561. A numerical program was
developed to analyse the properties of the lensed system, the results
of which can be summarized as follows. While the region of blueshifted
line emission is found to lie in between the two caustics, and is
therefore deflected into two images, redshifted emission is found
outside the caustics, and therefore results in a single lensed
image. We further argue that both redshifted and blueshifted line
emission originate from the same system: a disk with a molecular gas
mass of $\simeq 1-5\times10^{10}$M$_\odot$ in the host galaxy in
Q0957+561. Our results for the host galaxy are in excellent agreement
with a previous optical work by Keeton et al.\ (2000). To our
knowledge HR10 (Andreani et al.\ \cite{and}) and Q0957+561 are the
only systems at redshifts of $1<z<2$ in which CO emission was clearly
detected, but in contrast to Q0957+561 where we have not yet been able
to detect dust emission, HR10 is rich in dust and molecular
gas. Although both systems are significantly different, they both
independently give crucial insights into an epoch during which the
bulk of stars of the present day Universe formed, and thus mark an
important phase in galaxy evolution.

\begin{acknowledgements}
We are grateful to the referee Jean-Paul Kneib for his valuable
comments which have helped to improve the paper. We thank the PdB
staff for help with the data reduction. MK acknowledges funding
support by SFB grant 494. JMP has been partially supported by the
Ministerio de Ciencia y Tecnologia under grant numbers ESP2002-01627
and AYA2002-10113-E. PP acknowledges partial support by the Spanish
Ministerio de Ciencia y Tecnologia under grants ESP2002-01693,
ESP2003-04957 and AYA2003-07584.

\end{acknowledgements}

\end{document}